\documentclass[twocolumn,preprint]{article}
\usepackage{graphicx,amsmath,amsxtra,amssymb}
\begin{document}
\title{\Large \bf Pattern dynamics in a two-dimensional gas discharge system : \rm \\
  \bf Intermittent collapse and recreation, moving of spots,  dynamic change of clusters \rm }
\author{Takeshi Sugawara , Kunihiko Kaneko \\ \em Department of Basic Science, Graduate School of Arts and Science, \\ \em The University of Tokyo, 3-8-1, Komaba, Meguro, Tokyo 153-8902, Japan }

\date{}
\maketitle
\paragraph{Abstract}
\footnotesize
Reaction-diffusion equation model for a 2-dimensional gas discharge system is introduced in relationship with the recent experiment by Nasuno \cite{Nasuno's experiment}\cite{Nasuno's experiment2}.
The model shows formation of spots, molecule-like organization of a cluster of spots, moving of spots, and intermittent collapse and recreation of spots, in agreement with experiments.
The pattern dynamics are classified into distributed, localized, and moving spots, and periodic wave pattern.
The phase diagram displayed against the current and pressure has some agreement with that observed in the experiment.
Dynamic change of spot numbers as well as the structure of the spot cluster is studied in terms of the collective variable of charge, and discussed as chaotic itinerancy \cite{sugawara PTP}.
\normalsize
\section{Introduction}
\label{intro}
Pattern dynamics in non-equilibrium systems have been studied over decades, in fluid, chemical reaction-diffusion systems, liquid crystal, and so forth.
The extensive studies in the field have elucidated a rich variety of pattern dynamics, together with the advances in theoretical analysis \cite{mikhailov 3} $\sim$ \cite{pearson}.

One field in the pattern dynamics that is not so well explored in comparison with the above examples, is a discharge system.
When a strong voltage is applied between the electrodes, electric discharge appears through the gas filled in the chamber.
The discharge often forms a complex spatiotemporal pattern, as has been studied theoretically and experimentally (\cite{2 layers model 1}$\sim$\cite{Germany group 3}).

In such discharge system with a variable resister, there exists some global constraint among each local discharge processes.
Viewed as pattern dynamics, this means existence of global coupling among local dynamical processes.
On the other hand, global coupling among nonlinear units often shows a non-trivial collective motion, as has been extensively studied as collective motion in globally coupled dynamical systems.
Now in the discharge system, there exists interplay between collective motion in globally coupled systems and pattern dynamics by local nonlinear dynamics.
As a study of nonlinear system, it is interesting to search for some novel dynamic state, due to this interplay.

Indeed, there is a beautiful experiment by Nasuno, suggesting such novel, non-trivial dynamics\cite{Nasuno's experiment}\cite{Nasuno's experiment2}.
He set up an experiment consisting of two parallel plates between which electric charge flows.
By controlling the voltage and current, he found formation of spots, organization of 'molecular-like' structures of spots, complex motions of them, including the dynamics what he called teleportation.

In the present paper we introduce a phenomenological model on 2-dimensional (2D) gas discharge system, as a coupled electric circuit subjecting a global constraint, to describe the salient features observed in Nasuno's experiment, and to make further predictions.
As a theoretical model, we adopt a two-layer model that describes well a glow discharge system \cite{2 layers model 1}$\sim$ \cite{gas pressure effect 2}.
This two-layer model consists of a non-linear resistive part and a linear resistive part.
By extending this two-layer model, we construct a coupled dynamical system consisting of elementary circuit systems, each of which represents the process at the interior between square electrodes.
By taking spatial continuum limit, this model is reduced to a two-dimensional reaction diffusion (RD) system with global constraint.
Through extensive simulations of the model, we will show several phases of the pattern dynamics, and reproduce some salient features in Nasuno's experiment.

On the one hand, our model is an abstract and simplified model, and it may not completely correspond to the discharge system by Nasuno.
On the other hand, the model shows several novel interesting patten dynamics as a reaction-diffusion system with global coupling, which itself deserves investigation. 

The outline of the paper is as follows.
In Sec.\ref{Experiment}, the experiment by Nasuno is briefly described.
In Sec.\ref{Model}, we present some assumptions in order to reproduce the experiment and then construct a coupled circuit system as a model on the gas discharge experiment.
In Sec.\ref{Phenomena}, results of extensive simulations are presented, with classification of distinctive pattern dynamics as in the phase diagram.
Formation of spots, and a cluster of spots are numerically found, as well as the dynamics of creation and annihilation of spots, while the dynamical systems mechanism of the process is discussed.
In Sec.\ref{S and D}, summary, discussion and future perspectives are presented.

\begin{figure}[htbp]
\begin{center}
\includegraphics[scale=0.53]{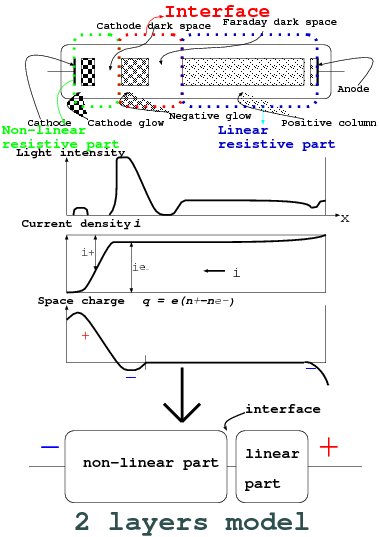}
\caption{Schematic figure of 2 layers model for glow discharge.
We consider that the glow discharge structure consists of two parts, the non-linear resistive region and the linear one.}
\label{glow discharge}
\end{center}
\end{figure}

\section{Experiment}
\label{Experiment}

Here we briefly describe a remarkable experiment by Nasuno \cite{Nasuno's experiment},\cite{Nasuno's experiment2}.
In the experiment, two square electrodes are immersed in low pressure air $P$ and connected to a dc power source $\epsilon$ and a variable resistor $R_{ex}$.  The power source $\epsilon$ can supply either a constant voltage of up to 2 kV or a constant current of up to 150 mA via a series resistor $Rex$ 35.2 k$\Omega$.

He adopted the parameter region of the gas discharge experiment so that the product of pressure $P$ and distance $d$ between electrodes satisfies $P \times d \sim$0.04 (Torr$\cdot$cm), by using $Air$.
The parameter region is located in the vicinity of the minimum of paschen curve, according to Fig.7.3 of chapter7 in \cite{discharge physics text}.
It is known that, under the experimental condition, current-voltage characteristics has monotonous curve \cite{discharge physics text}$\sim$\cite{negative slope problems 2}, and that no pattern dynamics is usually observed.
He fixed the total current $I$ instead of voltage, which is different from the conventional discharge experiments.
However, when the current $I$ is fixed as a constant,  $V$ is determined uniquely by a monotonous current-voltage characteristic so that $V$ is also constant. 
Indeed, according to the experiment, current-voltage characteristic $I-V$ is monotonous, so that $I$ and $V$ remain fixed\cite{Nasuno's experiment2}.  
However, local current and charge density between electrodes can change in space and time, to produce nontrivial dynamics.

\paragraph{}
By controlling $I$ and $P$, complex pattern dynamics in glow discharge region was observed.
By increasing current $I$, the pattern dynamics of discharge changes as follows.

\begin{itemize}
\item Just after onset of glow discharge, one isolated spot with a typical size (that is, a humped bell-like light intensity distribution) appears.
The spot wanders irregularly on the square plane.

\item With the increase of $I$, the number of isolated and wandering spots increases.

\item As $I$ is increased further, some spots form a molecule-like localized structure, which is called \em cluster\rm.
This cluster also wanders through the plane.

\item With the further increase of $I$, the number of spots in the cluster increases leading to various clusters with a variety of configurations of spots.
Switching among different clusters occurs intermittently.

\item At much higher $I$, the excited domain (the area with high light intensity) forms a string or a closed loop.

\item A remarkable phenomenon which he termed 'Teleportation' was reported,  where annihilation of a spot and immediate recreation of a different spot at a distant place was observed.
For example, when 3 spots existed, one of them suddenly was disappeared, and then a new spot immediately appeared at a distant position, located in the neighbor of a remaining spot, so that the 3 spot-state is recovered \cite{Nasuno's experiment}.

\end{itemize}

The phase diagram for the above behaviors is displayed in terms of $I$ and $P$ in the paper \cite{Nasuno's experiment}.

\section{Model}
\label{Model}
\subsection{Assumption}

Here we introduce an electric circuit model corresponding to the above experiment. 
The model is essentially based on a circuit model introduced by Purwins et al.\cite{2 layers model 3}$\sim$\cite{gas pressure effect 2}, derived from an analogy between an electric circuit and a reduced equation from plasma physics.
The model is at a macroscopic phenomenological level, where discharge process between 2D electrodes is described as \em charge transfer process by a coupled circuit system \rm.
For each element, we adopt the so-called ``2 layer model'' which consists of non-linear resistive part and linear resistive one, as schematically shown in Fig.\ref{glow discharge}.
The model has succeeded in explaining several discharge experiments\cite{2 layers model 1}$\sim$\cite{gas pressure effect 2}.
Here we revise their model so that it meets with the experimental condition by Nasuno.

To set up the coupled circuit model, the following assumptions are made following Purwins et al.

\begin{itemize}

\item The experimental condition is set at the region of glow discharge.

\item We describe the phenomena in terms of current density $i(x,y)$ and space charge density $q(x,y)$ because we focus only on the \em charge transfer process \rm between two electrodes.

\item \em Gas pressure effect \rm between electrodes known experimentally is taken into account.
According to \cite{gas pressure effect 1}\cite{gas pressure effect 2}, this effect is effectively described as diffusion of current density that flows between electrodes.
Based on the experimental results, we assume a monotonous relation between gas pressure and the diffusion of current density.

\end{itemize}

In considering Nasuno's experiment, we make further assumptions:

\begin{itemize}

\item \em Global \rm current-voltage characteristic shows a \em monotonous \rm cubic curve in a logarithmic scale \cite{footnote 2}, as will be given later(Fig.\ref{nullcline}(b)).
\item The total current that flows into electrodes is kept as \em constant \rm.
Therefore, the voltage drop between electrodes also remains constant.
\item In electrodes (represented by 2-dimensional planes), local voltage-current density characteristic is assumed to be cubic in a logarithmic scale with a \em negative slope. \rm

\end{itemize}

\subsection{Specific Model}

\begin{figure}[htbp]
\begin{center}
\includegraphics[scale=0.55]{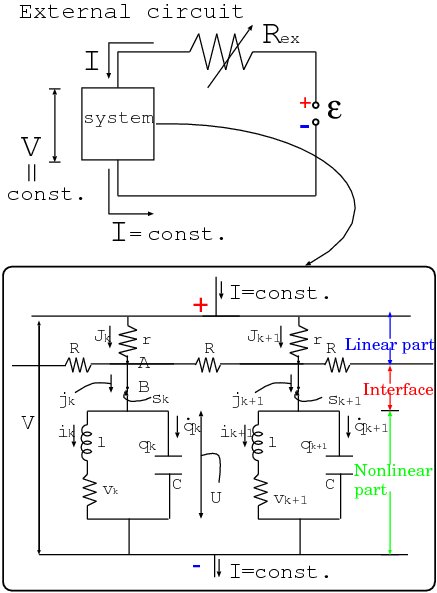}
\caption{Schematic figure of our 2-dimensional discharge model.
Our model system consists of $N$ coupled circuits.
The element $k$ is connected with neighbor ones by linear resistor $R$ and with ``External circuit''.
Although displayed in a one-dimensional representation, what we actually simulated in the paper is a two-dimensional case. }
\label{circuit model}
\end{center}
\end{figure}

The first set of assumptions underlie the coupled circuit model by Purwins et al.\cite{2 layers model 3}$\sim$\cite{gas pressure effect 2}, while we arrange it to meet with the latter set of assumptions to adapt to Nasuno's experiment.
\paragraph{}
The discharge model which we consider here consists of two square electrodes (area $S$) and the interior, which are under ``External circuit'' that consists of the dc power source $\epsilon$ and the variable resistor $R_{ex}$, as shown in Fig.\ref{circuit model}.
The total current $I$ flowing into the electrodes is controlled to be constant\cite{Nasuno's experiment}.
Thus $V$ is uniquely determined to be a constant.
Further, each elementary circuit $k$ is connected with the neighboring ones by linear resistor $R$ and is also contacted with ``External circuit''.
As shown in Fig.\ref{circuit model}, this elementary circuit consists of a linear resistor, capacitance, linear coil and a nonlinear resistor given by a specific $i$-$v$ characteristic.
The linear resistor $r$, the region U and the intermediate region correspond to linear resistive layer, the non-linear resistive layer, and the interface in 2-layer model, respectively.
Each part is characterized by a linear resistor $r$, capacitance C, the constant self-inductance $l$ of the coil, while the voltage drop $v_{k}$ at the non-linear layer follows the $i$-$v$ characteristic given by a log-scaled cubic function of $i_{k}$.
The total current $I$ is distributed to each element with a current $J_k$, while its current at B and U is given by $j_{k}$ and $i_{k} + \dot{q_{k}}$ respectively, as shown in Fig.\ref{circuit model}.

Now, the main discharge system is described by a set of $N$ ordinary differential equations as $N$ coupled circuits, as shown in Fig \ref{circuit model}.
In the model it is assumed that there are some charge flow into the interface (i.e. at "B" in the figure) from the non-linear resistive layer. 
This flow of charge density is represented \em by \rm $\dot{s_{k}}$.
The voltage drop at the interface is represented by $\frac{s_{k}}{C}$
Since the total space charge $Q$ should be constant in time between electrodes, there is a global constraint on $\sum_k s_k$ as will be discussed again in 
Sec.\ref{Flowing charge density}.

Considering the law of the conservation of charge (the first Kirchoff's rule) at the point A, we get
\begin{eqnarray}
\label{model-1}
J_{k}-j_{k}+[-\frac{1}{RC}\{(q_{k}+s_{k})-(q_{k-1}+s_{k-1}) \\
\nonumber +(q_{k}+s_{k})-(q_{k+1}+s_{k+1})\}]=0,
\end{eqnarray}
while the law of the conservation of charge at the point B gives
\begin{equation}
\label{model-2}
j_{k}-(i_{k}+\frac{dq_{k}}{dt})=\frac{ds_{k}}{dt}.
\end{equation}
At the region U, the voltage balance (the second Kirchoff's rule) leads to the following equation
\begin{equation}
\label{model-3}
l\frac{di_{k}}{dt}+v_{k}-\frac{q_{k}}{C}=0.
\end{equation}

In addition to (\ref{model-1})$\sim$(\ref{model-3}), we need to consider the constraint on the constant total current $I$ and constant voltage drop $V$.
\begin{gather}
\label{model-I}
I=\sum_{k=1}^{N}J_{k}=const. \\
\label{model-V}
V=rJ_{k}+\frac{q_{k}}{C}+\frac{s_{k}}{C}=const. 
\end{gather}

Here, to solve the above equations, we assume a constraint for the description of $s_{k}$ as a function of $\{i_{m},q_{m}\},(m=1,2,...,N)$.
Accordingly we must define $s_{k}$ satisfying
(\ref{model-1})$\sim$(\ref{model-V}).
After some calculations, with $v_{k}=v(i_{k})$ and $q^{\prime}_{k}=q_{k}+s_{k}$, $s_{k}$ are determined so as to satisfy
\begin{gather}
\frac{dq^{\prime}_{k}}{dt}=\frac{V}{r}-\frac{q^{\prime}_{k}}{rC}-i_{k}+\frac{1}{RC}(q^{\prime}_{k-1}+q^{\prime}_{k+1}-2q^{\prime}_{k})  \tag*{}
\\
\frac{di_{k}}{dt}=\frac{q^{\prime}_{k}}{lC}-\frac{s_{k}}{lC}-\frac{v(i_{k})}{l} \tag*{}
\\
I=\sum_{k=1}^{N}i_{k}=const. \tag*{} \\
Q^{\prime}=\sum_{k=1}^{N}q^{\prime}_{k}=Cv(\frac{I}{N})N=const.\tag*{} \\
V=r\frac{I}{N}+\frac{1}{C}\frac{Q^{\prime}}{N}=const.\tag*{}
\end{gather}
We define $s_{k}$ as
\[s_{k}\equiv\frac{i_{k}}{I}(Q^{\prime}-C\sum_{k=1}^{N}v(i_{k})).\]
(See appendix for details.)
By taking spatial continuum limit, we obtain the following 2-component RD equation, with
$D_{q^{\prime}}= \frac{1}{RC}, Q^{\prime}=\int q^{\prime} dS$, as 
\begin{subequations}
\begin{eqnarray}
\label{model-4}
\frac{dq^{\prime}}{dt}=\frac{V}{r}-\frac{q^{\prime}}{rC}-i+D_{q^{\prime}}\triangle q^{\prime} \\
\label{model-5}
\frac{di}{dt}=\frac{q^{\prime}}{lC}-\frac{s}{lC}-\frac{v(i)}{l}+D_{i}\triangle i \\
\label{model-6}
s\equiv\frac{i}{I}\{Q^{\prime}-C\int v(i) dS\} \\
\label{model-7}
I=\int i dS =const. \hspace{.2in} \\
\label{model-8}
Q^{\prime}=\int q^{\prime} dS =Cv(\frac{I}{S})S=const. \\
\label{model-9}
V=r\frac{I}{S}+v(\frac{I}{S})=const.
\end{eqnarray}
\end{subequations}

By using dimensionless variables 
\paragraph{}
$\tilde{i}=\frac{i}{i_{c}}$, $\tilde{v}=\frac{v}{v_{c}}$,
$\tilde{q}=\frac{q^{\prime}}{Cv_{c}}$, $\tilde{t}=\frac{t}{\frac{l}{r}}$
\hspace{.1in}and redefining

$\tilde{i}\rightarrow i$, $\tilde{v}\rightarrow v$,
$\tilde{q}\rightarrow q$, $\tilde{t}\rightarrow t$,\hspace{.2in}

the RD equation (\ref{model-4})$\sim$(\ref{model-9}) is written as
\begin{subequations}
\begin{eqnarray}
\label{mujigen-1}
\tau\frac{dq}{dt}=V-q-ai+\triangle q  \\
\label{mujigen-2}
\frac{di}{dt}=\frac{1}{a}\{q-s-v(i)\}+D\triangle i  \\
\label{mujigen-3}
s\equiv\frac{i}{I}\{Q-\int v(i) dS\} \\
\label{mujigen-4}
I=\int i dS =const. \hspace{.2in} \\
\label{mujigen-5}
Q=\int q dS =v(\frac{I}{S})S=const. \\
\label{mujigen-6}
V=a\frac{I}{S}+v(\frac{I}{S})=const.,
\end{eqnarray}
\end{subequations}
\paragraph{}
where
$\tau=\frac{rC}{\frac{l}{r}},a=r\frac{i_{c}}{v_{c}},D^{\prime}=\frac{D_{i}}{D_{q}},D=\frac{D^{\prime}}{\tau},\\
\hspace{.32in}\xi^{2}=rCD_{q}$ (characteristic length),
$S=\frac{1}{\xi^{2}}$.

\subsection{Flowing charge density}
\label{Flowing charge density}

Here we make some remarks on the flowing charge $s_{k}$ at the point B in Fig \ref{circuit model}.
The point B corresponds to the interface of the 2-layer model, that is, the position between the ``Negative glow" and Faraday dark space, which exists for the steady glow discharge (Fig \ref{glow discharge}).
In this region, electrons flowing from Cathode layer (non-linear resistive layer) combine with positive ions flowing from ``Positive column", or anode (i.e., the linear resistive layer).
There, some complex processes occur through diffusion, excitation, ionization, and recombination of molecules.
Hence there is a longitudinal flow of charge, and the flowing charge $s_{k}$ at B is distributed.

For a homogeneous steady glow discharge, there is no charge in this region. 
Furthermore, $s_{k}$ at B is zero, as the initial state before the discharge.
Hence, if a spatially homogeneous state were stable, the flow term $s_{k}$ would remain to be 0. 
However, as will be shown, for almost all the parameter regions, such homogeneous state is unstable, where charge density $q$ is distributed inhomogeneously in the non-linear resistive layer.
In this case, flowing charge density $s$ is also distributed in the interface.
Hence we need to consider this term.

After taking a spatial continuum limit, the total flowing charge is given by the spatial integration,
\begin{equation}
h\equiv \int s dS (= Q-\int v(i) dS ),
\end{equation}
where $h$ means the total charge flowing from the non-linear resistive layer into the interface, which is a macroscopic dynamical variable characterizing the global charge transfer process between the non-linear layer and the interface. 

\paragraph{}
In Nasuno's experiment \cite{Nasuno's experiment},\cite{Nasuno's experiment2}, due to the short distance between electrodes, the Positive column does not appear.
 Hence the anode and the anode surface correspond to the linear part and the interface, respectively.
Since the anode has very small resistance in the experiment, $s$ is assumed to be distributed over the anode surface.

\subsection{Parameter setting}
Hereafter we study the behaviors of the present model given by the equations (\ref{mujigen-1})$\sim$(\ref{mujigen-6}), by controlling the parameters $I$ and $D$, which are the total current into the system, and a parameter for the gas pressure effect, respectively.
Hence the control parameters $I$ and $D$ correspond to those in the experiment, i.e., the total current and the pressure $P$, respectively.

 As initial conditions of $i(x,y)$ and $q(x,y)$, we mostly choose ``uniform state $i_{u}, q_{u}$" with small fluctuations $\eta(x,y)$ as a uniform random number over [-0.01,0.01].
Accordingly total current $I$ (= $i_{u}S$), total charge $Q (=\int q dS)$ and voltage $V$ are determined initially (See appendix).
The phenomena to be discussed is always observed from these initial conditions, \em i.e. \rm the pattern dynamics is attracted to a global attractor. 
Unless otherwise mentioned, the other parameters are fixed as 
\paragraph{}
$\tau=20, \xi^{2}=0.004 $ (characteristic length),$ a=400\frac{i_{c}}{v_{c}}=0.0108$,$ S=\frac{1}{\xi^{2}}=250$.
\paragraph{}
Numerical simulations are carried out by the mesh size $128\times128$.
Unless otherwise mentioned, we choose a periodic boundary condition.

\begin{figure}[htbp]
\begin{center}
\includegraphics[scale=0.7]{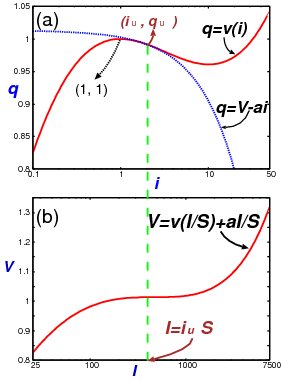}
\caption{Nullcline of the single element dynamics (a), and global $I-V$ characteristic (b).
We mainly focus on the case that the steady state ($i_{u},q_{u}$) is among the critical point (1,1), where it is stable if $i_{u} < 1$, whereas destabilized if $i_{u}> 1$ and then the element shows limit cycle oscillation.
$I-V$ characteristic is monotonous so that a steady state is realized.  
Q is also a function of $i_{u}$. 
$v(i)$ is cubic and log-scaled function of $i$. 
Note the relation $I=i_{u}S$.}
\label{nullcline}
\end{center}
\end{figure}

These parameters are chosen so that a single element satisfies the
behavior for a simple discharge (without spatial pattern).
Before describing the pattern dynamics of the model, we first show how a
single element behaves.
Fig \ref{nullcline} is the nullcline of the single element dynamics (Fig.\ref{nullcline}(a)), and global $I-V$ characteristic (Fig.\ref{nullcline}(b)).
We mainly focus on the case that the steady state ($i_{u},q_{u}$) is among the critical point (1,1) in Fig.\ref{nullcline}(a) \cite{icvc}.

\section{Phenomena}
\label{Phenomena}

\subsection{Global phase diagram}

According to our interpretation of the present model (\ref{mujigen-1})$\sim$(\ref{mujigen-6}), the discharge occurs at local site if $i(x,y) \geq i_{u}$. 
Hence we show the pattern dynamics by displaying only the pixels that satisfy $i(x,y)>i_{u}$, an active region with discharge, by gray.

\begin{figure}
\begin{center}
\includegraphics[scale=0.45]{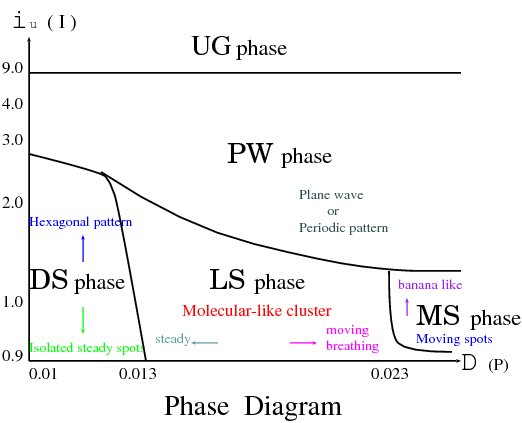}
\caption{Phase diagram as function of $i_{u} (=\frac{I}{S})$ and $D$.
Vertical axis is log-scaled.
Global property of the model is roughly classified into five phases under $i_{u}> 0.9 $.}
\label{global phase diagram}
\end{center}
\end{figure}

Here we study the pattern dynamics by changing $i_{u} (=\frac{I}{S})$ and $D$, by focusing on the parameter region with $i_{u} \sim 1$, where the discharge phenomena occur.
As a 2-dimensional phase diagram with regard to $i_{u}$ and $D$, the pattern dynamics we have observed is classified into five phases, as displayed in Fig.\ref{global phase diagram}.
We first give a brief description of each phase, and will discuss the characteristic of each phase in detail later.

For small $i_{u}(\sim 1.0)$, there are three phases, depending on the value of $D$, that are "distributed spot phase (DS)", "localized spots phase (LS)", and moving spot phase (MS), respectively, as shown in Fig \ref{global phase diagram}.
In these phases, spots are formed.
Discharge occurs locally, only within these spots.
Indeed, when $i(x,y)$ is plotted in space and time, it shows a stepwise increase at the border of each spot, and shows a high plateau within each spot.
The pattern dynamics here will be discussed from the configuration of spot patterns and their motion, with which the three phases are classified.
For all these three phases, the number of spots increases with $i_u$.
For a larger value of $i_{u}$, there appears a periodic pattern in space, which we call Periodic Wave (PW) phase.
For a further large value of $i_u$, there is a uniform glow (UG), without spatial inhomogeneity.
The behavior of each phase is summarized as follows;

\begin{itemize}
\item DS phase --- Spots are arranged with some distance on the average, which decreases with the increase of $i_u$.
For small $i_u$, a few spots are isolated with some distance, while hexagonal pattern of spots appears as $i_{u}$ is increased.
After transient time, these spots are fixed, and do not move.

\item LS phase ---
Spots form several clusters similar to molecular structures, as reported in the experiment by Nasuno.
The clusters show a dynamic, complex behavior with the formation, collapse, and regeneration.
\item MS phase --- A spot can move by itself in the space, which shows a soliton-like behavior without collapse.
\item PW phase --- There appears a periodic wave pattern in space and time.
\item UG phase --- There is a uniform glow, without spatial inhomogeneity.
\end{itemize}

\paragraph{Collective variable -- $\int{s}dS$ ($\equiv h$) --}
As a global measure characterizing the pattern dynamics, the integration of $s$, $h \equiv \int{s}dS$ is often useful.
In the subsequent subsections, the pattern dynamics will be characterized by the motion of $h$, which is a collective variable.
Now we discuss the pattern dynamics, by referring to the change of this collective variable $h$.
(Note that $h \ll Q$ in the present parameter setting.)

\subsection{DS phase}
\paragraph{Isolated steady spots}

Starting with initial condition $i(x,y)=i_{u}+\eta(x,y)$, competition for current among elements occurs, resulting in instability of the homogeneous state.
This leads to the formation of spots with localized high current.
These spots are of the same size.
Initially, each spot moves very slowly, and is arranged so that each is located with an equal distance, until a regular, stationary spot pattern is formed.

\begin{figure}[htbp]
\begin{center}
\includegraphics[scale=0.35]{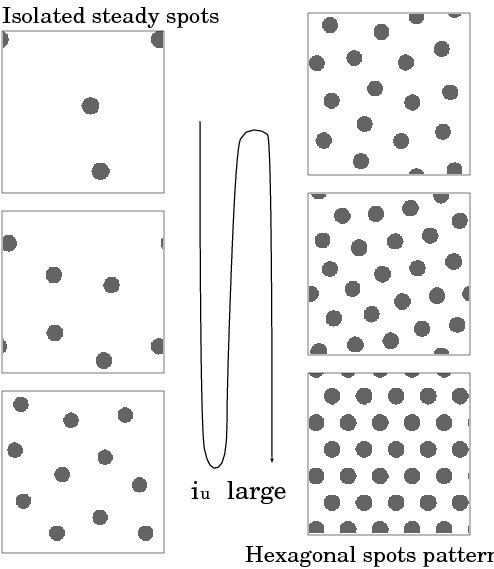}
\caption{Steady patterns observed in the DS phase.
Spots are of the same size under fixed $D$.
Initially, each spot moves very slowly, and is arranged so that it is located with the equal distance.
As $i_{u}$ is increased, the number of spots is increased.
As the number is increased with the increase of $i_u$, the distance between spots is decreased.
Left--Isolated steady spots.
$i_{u}=$0.87,0.93,1.0 in the order from the top to bottom with, $D$=0.01.
Right--Hexagonal spots pattern.
$i_{u}=$1.2,1.5,2.0 in the order from the top to bottom with $D$=0.01.
The mesh size here is 256$\times$256 }
\label{alpha-phase}
\end{center}
\end{figure}

\paragraph{Hexagonal pattern}
As $i_{u}$ is increased, the number of spots is increased, so that the distance between spots is decreased.
As $i_u$ is increased from $1.5$ to $2.0$, a regular hexagonal pattern of spots is formed, as shown in Fig.\ref{alpha-phase}.
The formation of this regular lattice of spots is understood through weak repulsion between neighboring spots.

\paragraph{Behaviors at the boundary between DS and other phases}

\begin{figure}
\begin{center}
\includegraphics[scale=0.47]{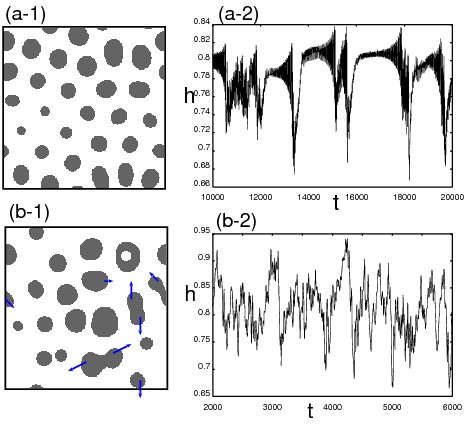}
\caption{Behaviors at the boundary between the DS and PW phases(a) and at the boundary between the DS and LS phases (b).
The left figures (1) give snapshots of pattern dynamics, while the time series of the integration of flowing charge $h(=\int{s}dS)$ corresponding to them are plotted in (2)(right).
In (a), the spots start to breathe, so that their sizes change, while the distance still remains almost equal.
In (b), the distance is no longer equal, and an irregular pattern is formed.\hspace{.2in}(a) $i_{u}=2.4,D=0.01$,\hspace{.1in} (b) $i_{u}=2.0,D=0.012$, mesh size 256$\times$256 }
\label{intermittency}
\end{center}
\end{figure}

As $i_{u}$ is increased toward the boundary to the PW phase, the spots start to breathe, and their sizes (and shape) change, while the distance between spots still remains almost equal as in Fig.\ref{intermittency}(a-1).
The collective variable $h$ changes intermittently, corresponding to irregular pattern dynamics (Fig.\ref{intermittency}(a-2)).

Around the boundary between the DS and LS phases, the characteristic wavelength disappears, and the configuration of spots is irregular, while spots start to show a bursting behavior, as well as splitting, as in (Fig \ref{intermittency}(b-1)).
The change of $h$ as well as the pattern dynamics is irregular(Fig \ref{intermittency}(b-2)).

\paragraph{}

In the experiment by Nasuno, the behavior corresponding to this phase is not observed.
We expect that the range of the change of pressure in the experiment may not be sufficient to cover low values needed and to detect this phase.

\subsection{LS phase}
\label{betaphase}

For small $i_{u}$($\sim$ 0.9), a single spot appears.
As $i_{u}$ is increased, the number of spots increases, through successive split of the original spot.
Spots are not completely stable in time, and some of them collapse after breathing.
As replication and extinction of spots are repeated, a cluster of spots is formed.
For small $i_u$ , the number of spots changes between 1 and 2, while the range of the number of spots that the system can take increases with $i_u$.
Some of the generation processes are displayed in Fig \ref{split}.

\paragraph{Phase diagram within the LS phase}

\begin{figure}[htbp]
\begin{center}
\includegraphics[scale=0.55]{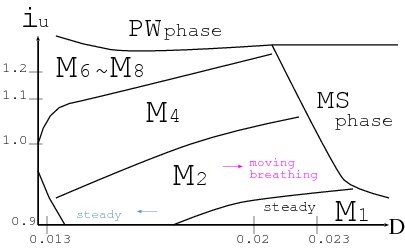}
\caption{Phase diagram of the LS phase.
The behaviors in the LS phase are classified in terms of the maximal number of spots, and plotted as a phase diagram in the $D-i_{u}$ parameter space.
Note that, in this phase diagram, ``$M_{n}$'' shows the case in which the maximal number of spots is $n$.}
\label{beta phase diagram}
\end{center}
\end{figure}

By measuring the dependence of the maximal number of spots on $i_u$ and $D$, the phase diagram within the LS phase is depicted as shown in Fig.\ref{beta phase diagram}, where the maximal number of spots is plotted in the $D-i_{u}$ parameter space.
Hereafter denote  \em the case ``$M_{k}$'' \rm when $k$ spots exist at maximum.
On the other hand, if $n$ spots exist at a moment, we call \em the state ``$sp_{n}$''\rm .
Indeed, the configuration of the phases in the diagram of Fig.\ref{beta phase diagram} obtained from our model agree rather well with that obtained experimentally by Nasuno\cite{Nasuno's experiment}.

\paragraph{Localized cluster of spots; molecule-like structure}

In the following, we discuss dynamics of a localized cluster of spots.
After spots are generated by splitting, they separate up to some distance.
Thus they form a chain-like structure, a localized structure like a molecule.
As the distance between spots is increased and a spot is separated from a cluster very slowly, it starts breathing.
Following the amplification of this oscillation, the spot collapses.
After this collapse, split of some other spot(s) follows, and the number of spots returns to the maximal under the given value of $i_{u}$.
In this way, a cluster of spots is sustained.

\begin{figure}[htbp]
\begin{center}
\includegraphics[scale=0.55]{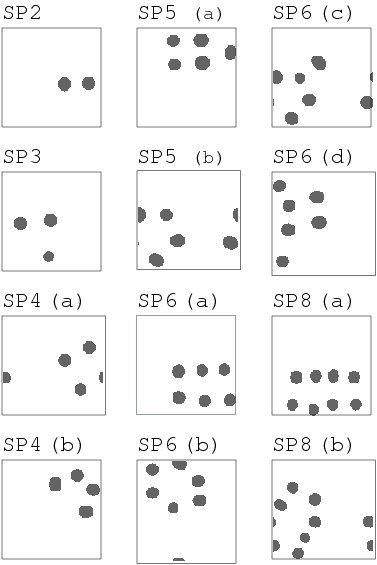}
\caption{Snapshots of typical clusters,$sp_{2}\sim sp_{8}$.
They are the snapshots of the clusters with the maximal number of spots under the given $i_{u}$ and $D$.
$D$=$0.014.\hspace{.03in}sp_{2}i_{u}$=$0.93$,$sp_{3}i_{u}$=$0.934$,$sp_{4}$(a)$i_{u}$=$0.98,\hspace{.02in}
 \newline sp_{4}$(b)$i_{u}$=$1.01,\hspace{.02in}sp_{5}$(a)$i_{u}$=$1.1$,
$sp_{5}$(b)$i_{u}$=$1.06, sp_{6}$(a)
\newline $i_{u}$=$1.1,\hspace{.02in}sp_{6}$(b)$i_{u}$=$1.1$,$sp_{6}$(c)$i_{u}$=$1.1,\hspace{.02in}sp_{6}$(d)$i_{u}$=$1.1,\hspace{.02in}
\newline sp_{8}$(a)$i_{u}$=$1.15$,$sp_{6}$(b)$i_{u}$=$1.15$.}
\label{localized structures}
\end{center}
\end{figure}

\begin{figure}[htbp]
\begin{center}
\includegraphics[scale=0.7]{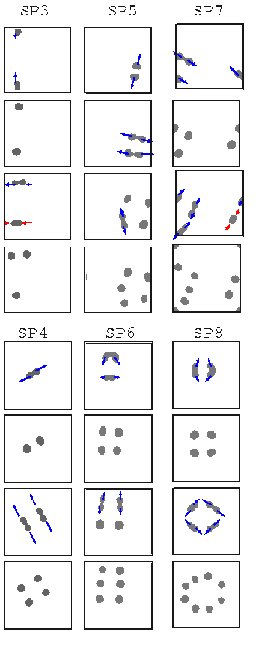}
\caption{Generation processes of the clusters $sp_{3} \sim sp_{8}$ through successive split of the spots.
$sp_{3}$ : $sp_{1} \rightarrow sp_{2} \Rightarrow sp_{2}\rightarrow
sp_{3} \Rightarrow sp_{3}$ cluster ($i_{u}=0.939,D=0.014$).
$sp_{4}$ : $sp_{1} \rightarrow sp_{2} \Rightarrow sp_{2} \rightarrow
sp_{4} \Rightarrow sp_{4}$ cluster ($i_{u}=0.98,D=0.014$).
$sp_{5}$ : $sp_{2} \rightarrow sp_{4} \Rightarrow sp_{4} \rightarrow
sp_{5} \Rightarrow sp_{5}$ cluster ($i_{u}=1.06,D=0.013$).
$sp_{6}$ : $sp_{2} \rightarrow sp_{4} \Rightarrow sp_{4} \rightarrow
sp_{6} \Rightarrow sp_{6}$ cluster ($i_{u}=1.08,D=0.013$).
$sp_{7}$ : $sp_{2} \rightarrow sp_{4} \Rightarrow sp_{4} \rightarrow
sp_{7} \Rightarrow sp_{7}$ cluster ($i_{u}=1.1,D=0.013$).
$sp_{8}$ : $sp_{2} \rightarrow sp_{4} \Rightarrow sp_{4} \rightarrow
sp_{8} \Rightarrow sp_{8}$ cluster ($i_{u}=1.1,D=0.013$).}
\label{split}
\end{center}
\end{figure}

When the maximal number of spots $\geq$ 6, there appears a variety of configurations for the cluster, as shown in Fig \ref{localized structures}.
Depending on which spot splits to which direction, there appears a different configuration.
Over a long time span, one can observe a variety of spot configurations.

\begin{figure}[htbp]
\begin{center}
\includegraphics[scale=0.6]{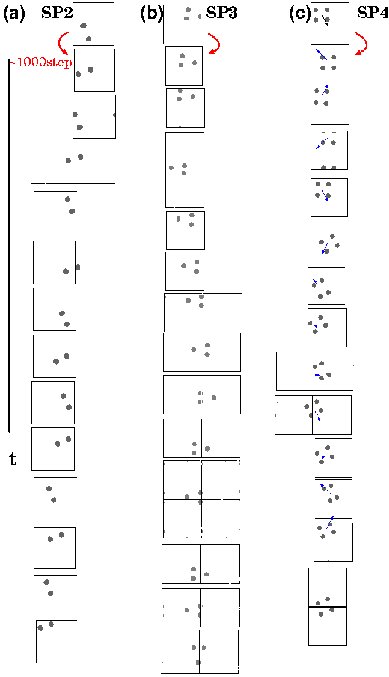}
\caption{A sequence of snapshots for $sp_{2},sp_{3},sp_{4}$ state.
We see that the positions of $sp_{2},sp_{3},sp_{4}$ shift in time due to the asymmetry for splitting.
In a long time scale, the clusters appear to wander.
(a) $sp_{2}$:$i_{u}$=$0.93$($M_{2}$case), (b) $sp_{3}$:$i_{u}$=$0.939$($M_{3}$case), (c) $sp_{4}$:$i_{u}$=$0.98$($M_{4}$case), $D$=$0.014$}
\label{sp2-snap-series}
\end{center}
\end{figure}

In a long time scale, a cluster wanders slowly on a 2-dimensional space, since the configuration after reconstruction of a cluster is slightly different from the original, even if they are similar.
After the split of a spot, there remains some asymmetry between the two spots, which causes a wandering motion of a cluster.
We show three examples of such wandering cluster, by displaying successive snapshot patterns in Fig. \ref{sp2-snap-series}, where the maximum number of spots is 2 ($M_{2}$) for Fig \ref{sp2-snap-series}(a), 3 ($M_{3}$) for Fig \ref{sp2-snap-series}(b) and 4 ($M_{4}$) for Fig, \ref{sp2-snap-series} (c), respectively.
The motion of cluster is not unidirectional, but its direction changes irregularly.
This irregular motion arises since the two spots after split are not completely identical.

\paragraph{Dynamics of the collective variable $h$ corresponding to the change in the spot number}

\begin{figure}[htbp]
\begin{center}
\includegraphics[scale=0.42]{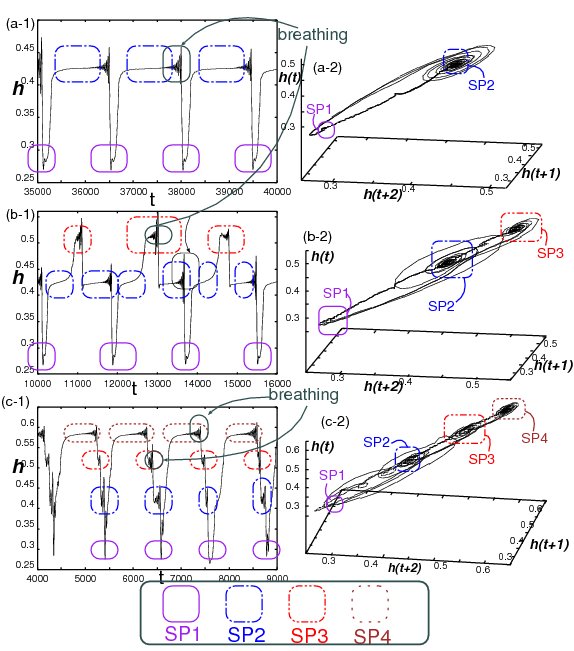}
\caption{Dynamics of $h$. (1) Its time series and (2) orbits embedded into a 3-dimensional phase.
(a) The case $M_{2}$ ($i_{u}$=0.93), (b) The case $M_{3}$ ($i_{u}$=0.939), (c) The case $M_{4}$ ($i_{u}$=0.98). 
Each plateau value of $h$ corresponds to each state $sp_{1}\sim sp_{4}$ with respect to different $i_{u}$ under fixed $D$.
$D=0.014$ }
\label{Time series of h for sp2 to sp4}
\end{center}
\end{figure}

When the number of spots changes, the transfer of charge is altered drastically. Hence, it is relevant to study the number change in relationship with the collective variable $h$, i.e., the integration of the flowing charge.
In Fig \ref{Time series of h for sp2 to sp4}(a),(b), and (c), we show the time series $h$ for $M_{2}$, $M_{3}$ and $M_{4}$ cases.
We plot the time series of $h$ (in (a-1),(b-1),(c-1)), as well as the orbit in the phase space, by embedding the time series of $h$ into three dimensional phase space ($h(t),h(t+1),h(t+2)$).

For all the cases, $h$ increases with the spot number $n$.
When the spot number stays at some value, $h$ remains almost constant.
Note that $h$ value is mainly determined by the spot number of the moment $sp_n$, and is almost independent of the maximal number $M_k$ for a given condition.
For example, $h$ for $sp_{2}$ under $M_{2}$ almost equals to that of $sp_{2}$ under $M_{4}$.
If the system stays at a state with given $sp_{n}$ ($n \leq N$), it is clearly seen as each plateau in the time series of $h$, or a region with residence of an orbit around a fixed point in the phase space.
With the breathing of spots, $h$ also starts to oscillate, and in the phase space picture, the orbit spirals out of a fixed point, leading to a switch to a state with a different number of spots.

Consider the case M3 or M4.
There the decrease in the number of spots occurs as $(sp_{4}\rightarrow ) sp_{3}\rightarrow sp_{2} \rightarrow sp_{1}$, successively.
During this decrease, the system stays at each state with an intermediate spot number, for some time interval, while the return process $sp_{1}\rightarrow sp_{2} \rightarrow sp_{3}$ (or $sp_{4}$), is relatively rapid, without staying long at each intermediate state.
(The state $sp_1$ is unstable, and the orbit exits immediately).
These decrease and increase in the spot number are repeated.
We also note that the switching process observed here is true for a state with a higher number of spots, as have been numerically confirmed up to $sp_{8}$.
In Fig.\ref{M6-M8}, we display an example for $M_{6}\sim M_{8}$ case.

In the three-dimensional representation of the phase space from the time series of $h$, each state of a given spot number seems to be regarded as a saddle which has one-dimensional stable manifold, and two-dimensional unstable manifold, with an unstable focus.
The attraction and repulsion of each orbit around a saddle appears whenever the orbit passes through a given $sp_{n}$ state.
Each state of a given spot number is approached from a certain direction, and the orbit spirals out from it.
This process is analogous to that observed in Shilnikov chaos\cite{Shilnikov chaos}.
In the Shilnikov chaos, however, the spiral motion is stable, giving a stable 2-dimensional manifold for a focus, while a third direction gives an unstable manifold.
In contrast, in the present case, the former is unstable, and the third direction gives a stable manifold.
It is expected that the instability here leads to irregular wandering of spots.

\begin{figure}[htbp]
\begin{center}
\includegraphics[scale=0.4]{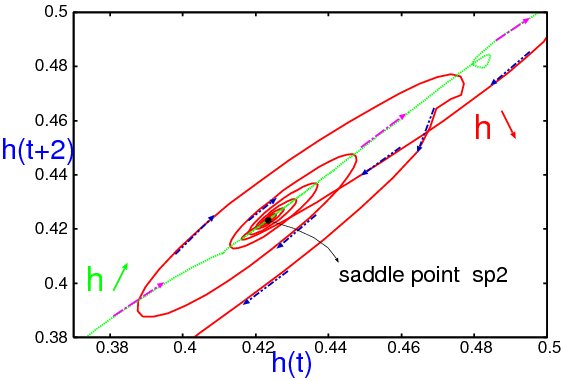}
\caption{Motion around a saddle in high dimensional phase space.
The nature of the saddle point $sp_{m}$ is different depending on if $h$ increases or decreases.  The approach to and deviation from a saddle  changes according to the in(de)crease of $h$.
$i_{u}=0.939,D=0.014$ ($M_{3}$ case, m=2) }
\label{saddle}
\end{center}
\end{figure}

\begin{figure}[htbp]
\begin{center}
\includegraphics[scale=0.42]{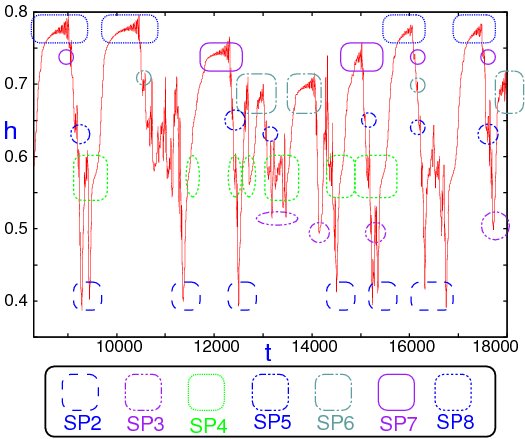}
\caption{Time series of $h$ for $i_{u}$=$1.1,D$=$0.013.$ 
$h$ shows complex dynamics, corresponding to the diversification of the cluster configurations with the increase of the spot number.}
\label{M6-M8}
\end{center}
\end{figure}

However, when $h$ increases, the orbits path through several saddle points with three real eigenvalues in this representation of the phase space, as long as  $sp_n$ is less than its maximal spot number (as can be seen in Fig.  \ref{Time series of h for sp2 to sp4}).
Note that  the orbits approaching the saddle and leaving it take different paths, depending on either if $h$ is increased or decreased, as shown in Fig.\ref{saddle}.  In other words, the three-dimensional phase-space  representation  is insufficient, and the high-dimensionality in the original dynamics leads to the dependence on the history of the change in $h$.
In this sense, the itinerant motion over different spot numbers will be better described as chaotic itinerancy\cite{GCM,chaotic scenario}, in the sense that several low dimensional ordered states are visited through high-dimensional chaotic motion.

\paragraph{Transition rules between several clusters}
Here we study transition-rules between clusters in some detail.

\begin{figure}[htbp]
\begin{center}
\includegraphics[scale=0.4]{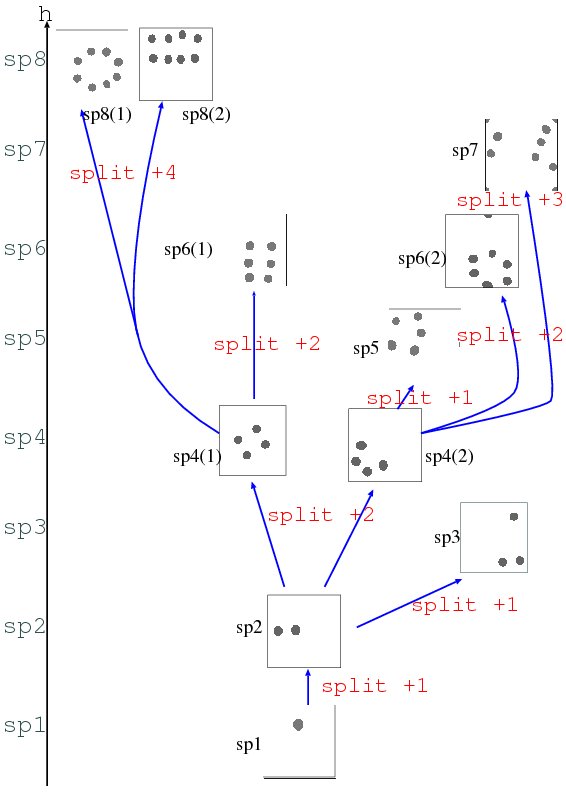}
\caption{Paths to create state with a higher number of spots, $sp_{n}$ (n=2,...,8) by splitting.
Note that only the states generated by splitting are displayed, and some other complicated ones, which are generated as a result of extinction of spots by breathing, are not displayed here.}
\label{keito}
\end{center}
\end{figure}

As the increase in the spot number occurs through the splitting of a spot, there exists a certain transition rule from a state with a lower number of spots to that with a higher number.
Here we display the paths of generating $sp_{n}(n=2,...,8)$ through the splitting process (see Fig.\ref{keito}).
As shown, each cluster with a larger number of spots is systematically generated from a \em specific \rm state with a smaller number.
Note that there are several states with different configuration of spots for a given number of them, when it is larger than or equal to 6 \cite{footnote 3}.

\begin{figure}[htbp]
\begin{center}
\includegraphics[scale=0.38]{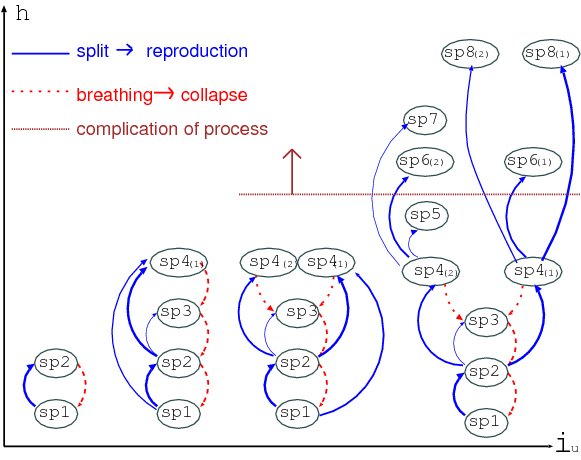}
\caption{Schematic figure for transitions among quasi-stationary states.
The solid upward and broken downward arrows show multiplication processes of spots by splitting and extinction process of them by breathing, respectively.
The thickness of the arrows indicates the frequency of the processes.
The typical cluster configurations in the present figure are shown in Fig \ref{keito}.}
\label{process}
\end{center}
\end{figure}

This transition rule of clusters is enriched with the increase of $i_{u}$.
In Fig.\ref{process}, Schematic figure for the change of the transition rule with $i_{u}$ is displayed, corresponding to the pattern in Fig \ref{keito}, where the solid upward and broken downward arrows show multiplication of spots by splitting and their collapse by breathing, respectively, while the thickness of the arrows shows the frequency of such processes observed.
The transition rules are summarized as follows.
\begin{itemize}
\item When $i_{u}$ is small, only the transition $sp_{1}\rightleftharpoons sp_{2}$ occurs.
\item As $i_{u}$ is increased, the states $sp_{3}$ and $sp_{4}^{(1)}$ (which is one type of the 4-spot state) appear, with the transitions $sp_{1}\sim sp_{3}$, $sp_{1}\sim sp_{4}^{(1)}$.
\item With the further increase of $i_{u}$, another 4-spot cluster with asymmetric configuration appears, denoted by $sp_{4}^{(2)}$.
Now there are transitions $sp_{1}\sim sp_{3}$, $sp_{1}\sim sp^{(1)}_{4}$, $sp_{1}\sim sp^{(2)}_{4}$.
\item As $i_{u}$ is increased further, the states $sp_{5}\sim sp_{8}$ appear, with a variety of transitions.
These transitions are divided into two groups as in Fig.\ref{keito}, that is, those among $sp_{1}\sim sp^{(1)}_{4}$ and among $sp_{1}\sim sp^{(2)}_{4}$.
\end{itemize}

Note that with the increase in the spot number, both the configurations and transitions are diversified, while $h$ shows complex dynamics corresponding to the diversification (Fig.\ref{M6-M8}).

\begin{itemize}
\item When the maximum number of spots is larger than or equal to 4, there exist clusters with the same spot number but different configurations.
Those states give almost the same value of $h$, and they are degenerated in the representation of $h$.
\item When the maximum number of spots is larger than or equal to 6, the collapse process is diversified, and indeed is more diverse than that displayed in Fig \ref{process}.
\item The higher the symmetry in the configuration of the clusters is, the higher is the frequency of the appearance of such configuration.
Indeed, this frequent appearance of configuration with higher symmetry is also observed in the experiment (\cite{Nasuno's experiment}\cite{Nasuno's experiment2}).
\item The transitions between the degenerate states, i.e., different configuration states with the same number of spots, cannot occur directly.
Only after the increase or decrease in the spot number, they can mutually change, as in Fig. \ref{process}.
\end{itemize}

\paragraph{Correspondence with the experiment}
\label{Correspondence with the experiment}

The pattern dynamics we observed here agree with that observed experimentally \cite{Nasuno's experiment}\cite{Nasuno's experiment2} as is summarized as follows.

\begin{itemize}
\item The global phase diagram on the spot number with regards to $i_u$ and $D$ (pressure) agree rather well.
\item The cluster of spots forms a molecule-like structure, whose structures are identical.
These structures are similar to that observed on the experiment.
(In the experiment, the molecular state is so far reported up to the spot number 6 mainly).
In our case, the number of spots shows intermittent decrease, while by comparing the state that is dominant in time, the agreement is clear.
\item When the maximal spot number is larger than or equal to 6, a variety of forms of clusters appear, both in our model and in experiment.
\item An interpretation of teleportation in the term of Nasuno: the reported phenomena in which one of the two spots disappears, and right after it, a new spot appears around the remaining spot.
Since it is observed as if one of the spots moved to a distant place within a very short time scale, he called this phenomenon as 'teleportation'.
According to our result, we can give a possible interpretation to this phenomenon.
In the $M_2$ regime, one of the spots disappears intermittently as we have mentioned.
Following this disappearance of one spot, the remaining spot divides into two.
Note that the time interval between the disappearance of one spot and the division of the remaining spot is very short.
Then, within the resolution of experimental measurement, this process is observed as if one spot teleportated to the location close to another spot.
\end{itemize}

\subsection{Moving spot phase}

At this phase a single spot moves in the space.
For $0.93\leq i_{u}\leq 1.0$ in Fig \ref{global phase diagram}, a single \em moving spot \rm appears, while multiple moving spots are observed with the increase of $i_{u}$.

\paragraph{Soliton-like spot}

\begin{figure}
\begin{center}
\includegraphics[scale=0.45]{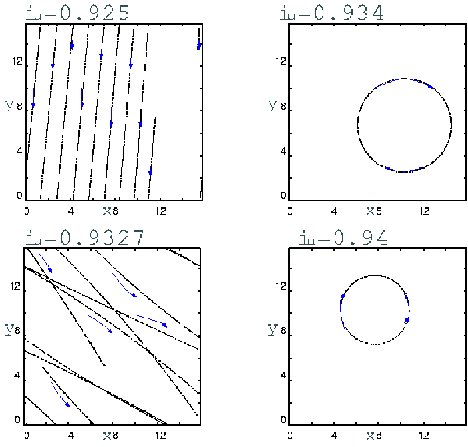}
\caption{Dependence of the orbits of a single moving spot on the value of $i_{u}$, under periodic boundary condition where a solid line shows the center of mass of the spot.
We show the direction of movement of ms1 with an arrow.
$D=$0.027, mesh size 256$\times$256.}
\label{Dependence under periodic condition}
\end{center}
\end{figure}

Here a single spot (abbreviated by \em ``$ms_{1}$'' \rm) moves without split or collapse.
The locus of a spot is displayed in Fig \ref{Dependence under periodic condition}, with a periodic boundary condition, where a solid line shows the center of mass of $ms_{1}$.
The center of mass is defined by the mean position among sites with $i(x,y)\geq i_{u}$.
Of course, if the spot is completely symmetric in shape, it cannot move.
The motion of a spot is due to asymmetry in the shape of $ms_{1}$, and indeed the spot is deviated from a circle slightly.

The nature of motion changes with the increase of $i_u$.
For small $i_u$ value, that is just above the onset of the appearance of $ms_{1}$, the motion is linear with a constant speed.
(See Fig \ref{Dependence under periodic condition},$i_{u}=0.925$).
With the increase of $i_{u}$, the motion starts to have a curvature, and the locus is bended.
This curvature is increased with $i_u$, and the locus shows a circle, as in Fig \ref{Dependence under periodic condition}, $i_{u}=0.934,0.94$.
With the increase of curvature, the speed also increases.

\begin{figure}
\begin{center}
\includegraphics[scale=0.5]{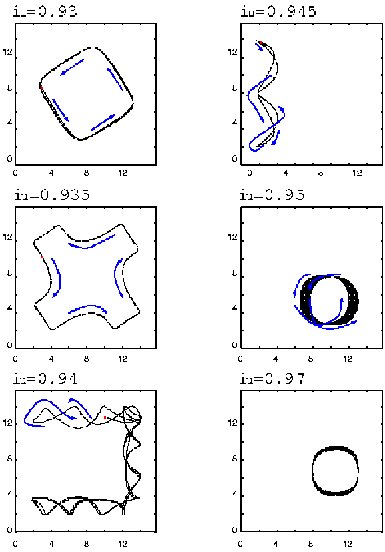}
\caption{The loci of a single moving spot under the Neumann boundary condition.
Arrows show the time course of the spot center.
Because the spot reflects at the boundary, the motion is complex as shown for $i_{u}=0.94$.
On the other hand, it often shows periodic orbits as given for the loci for $i_{u}=0.93,0.935,0.945.$
For $i_{u}=0.95,0.97$, it shows almost circular motion, but the center of the circular orbits shifts gradually.
These behaviors depend on the ratio of the radius of curvature of the spot locus to the electrode size $S^{\frac{1}{2}}$. $D=$0.027 }
\label{The orbits of center of mass on Neumann boundary condition}
\end{center}
\end{figure}

For this motion of spot, the choice of boundary condition may be more crucial. For example, to compare with an experiment, the periodic boundary condition may not be relevant.
Hence, we have also made some simulations with Neumann boundary condition, as shown in Fig \ref{The orbits of center of mass on Neumann boundary condition}. With reflection at the boundary, the motion becomes more complex.

\begin{figure}
\begin{center}
\includegraphics[scale=0.4]{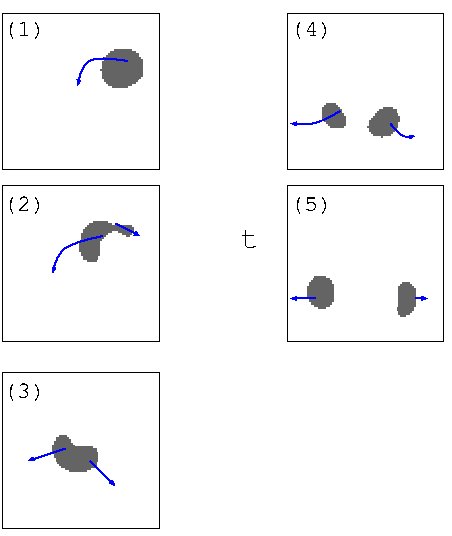}
\caption{Successive snapshots of multiple moving spots.
We show the direction of movement of the moving spots with arrows.
(1)A single moving spot just before splitting.
Here the size starts to expand.
(2)It splits into two spots, but one of them is too small and vanishes.
The other distorted spot survives and soon begins splitting again.
(3)Just before splitting of the distorted spot.
(4)Just after splitting of the distorted spot, leading to the state $ms_{2}$.
Each spot is separated from each other in the opposite direction.
(5)The state $ms_{2}$ is maintained, but one of them starts breathing and will vanish before long.
$i_{u}=1.05,D=0.03$ }
\label{ms2-snap}
\end{center}
\end{figure}

\paragraph{Multiple moving spots}
As $i_{u}$ is increased further, a single moving spot splits, leading to moving multiple spots(Fig \ref{ms2-snap}).
Here, the moving spot is first distorted to extend its tail as spiral.
When the tail is small, the spot rotates with it, while as it is larger, the spot is divided into two, which moves to the opposite direction (\em generation of $ms_{2}$\rm).
After the motion of these two spots, collapse of one spot occurs as in the LS phase, and the system comes back to a single spot state.
These processes are repeated to lead to a complex motion of spots.

\begin{figure}[htbp]
\begin{center}
\includegraphics[scale=0.5]{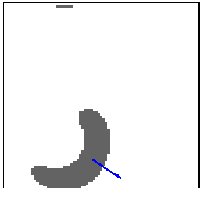}
\caption{Banana-like spot.
A spot is spread in the direction perpendicular to the direction of its motion, and finally the banana-like shape is formed.
The banana-like spot moves straight without collapse.
$i_{u}=1.2,D=0.03$ }
\label{banana-snap}
\end{center}
\end{figure}

As shown in Fig.\ref{ms2-snap}, the spot shape starts to be distorted.
This distortion is much clearer, as $i_u$ is further increased, where the system approaches the boundary with the PW phase.
There the spot size is larger, and forms a distorted "banana-like" shape (Fig.\ref{banana-snap}).
This "banana-like" spot moves straight without collapse.

\begin{figure}
\begin{center}
\includegraphics[scale=0.42]{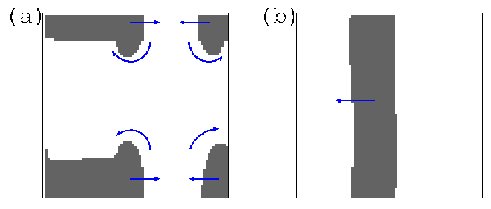}
\caption{ (a) Periodic pattern. 
The parameter values are $i_{u}=5.0$, $D=0.015$, while the similar patterns appear over a wide range of parameters.
(b) Plane wave. $i_{u}=5.0$, $D=0.03$.
This plane wave moves in one direction as shown by the arrow.}
\label{delta-phase}
\end{center}
\end{figure}

\paragraph{Periodic wave phase}
In general, as $i_{u}$ is increased, each site tends to synchronize with each other, since the effective (attractive) coupling among oscillator elements is stronger.
Accordingly, in this regime of a large current, spots no longer exist, and they are replaced by a pattern periodic both in space and time (Fig \ref{delta-phase}(a)).

\paragraph{Traveling wave}

One typical pattern observed here is plane wave(Fig \ref{delta-phase}(b)).
As shown, an extended string of discharged region propagates with a constant speed here.
A string of concentrated electric discharge is formed over the plate, and it travels with a constant speed.

\begin{figure}[htbp]
\begin{center}
\includegraphics[scale=0.4]{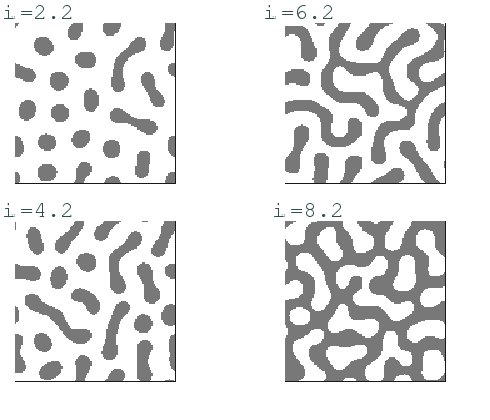}
\caption{Steady stripe patten.
Spots are extended and connected to form a \em string \rm ($i_{u}=2.2,4.2$).
As $i_{u}\nearrow$, many strings are formed, and as a result they are connected with the neighboring strings as shown (\em stripe\rm).
Plotted for $i_{u}=6.2,8.2$.}
\label{Stripe pattern}
\end{center}
\end{figure}

\paragraph{Stripe pattern}
Here we briefly describe pattern observed at a different set of parameter values of $\tau$ and $\xi$ than that adopted so far.
For some parameter values of $\tau$ and $\xi$, we have found a steady pattern, which is different from the DS phase.
As an example, we discuss briefly the pattern observed for $\tau=35$ and $\xi$=0.008.
As $i_u$ is increased, the connected spots form a string and form a stripe pattern as shown in Fig \ref{Stripe pattern}.
The length of string is increased with $i_{u}$, and finally a stripe pattern covers the whole space as in Fig.\ref{Stripe pattern}.
Through the time evolution, a steady pattern is formed, without temporal change.
Such pattern formation is also common to that observed in Benard convection with a large aspect ratio.

\section{Summary and Discussion}
\label{S and D}
\subsection{Summary}

To sum up we have introduced a coupled circuit model $(\ref{mujigen-1})\sim (\ref{mujigen-6})$, corresponding to the discharge experiment by Nasuno.
Mathematically, the model belongs to a class of reaction-diffusion system with global coupling due to the global constraint on the conservation of charge and current.
When the total current is large, the system shows a homogeneous glow, while this homogeneous state is destabilized with the decrease in the current, and a variety of pattern dynamics of discharge is observed, as are classified into four phases, depending on the value of total current and the diffusion constant that corresponds to the pressure value in the discharge experiment.

For the regime with a lower current, the regions with concentrated discharge form spots.
They are classified as
\begin{itemize}
\item
{\bf Distributed Spot phase}:
Spots are arranged with some distance, to form a regular array of spots.
The phase appears for small $D$.

\item
{\bf Localized Spot phase}:
A few number of local spots exist, which form a cluster.
With the increase in the current the spot number increases.
In the cluster spots are arranged like a molecule, while there are several configurations of spots when their number is larger than or equal to 4.
Also, we have observed intermittent collapse of spots to decrease their number, and immediate recovery by the splitting of spots.
These processes are repeated as a cycle, and with this cycle, the cluster of spots wanders throughout the space.

\item
{\bf Moving Spot phase}:

For a larger value of $D$, a shape of a spot is asymmetric, and each spot starts to move by itself.
With the increase of the total current the motion changes from linear to circular, while complex motion is observed when the boundary condition is not periodic.

\end{itemize}

Besides these spot phases, we have observed another phase at a high current region:

\begin{itemize}

\item {\bf Periodic Wave phase}, where a string of discharged region propagates in space.

\end{itemize}
These pattern dynamics are characterized by introducing the collective variable $h=\int s dS$, the integration of the flowing charge $s_{k}$, which expresses the global charge transfer between the non-linear resistive region and the interface in the direction of the gap.
Indeed, $h$ turns out to characterize the number of spots, and the birth-and-death dynamics of spots are represented by the temporal change of $h$.

\subsection{Comparison with the experiment by Nasuno}

The behaviors observed in the LS phase reproduce the phenomena observed in experiments.
They include:

\begin{itemize}

\item
Increase of spot number with the current.

\item
Molecule-like structures of spots and their variety in shapes.

\item
Disappearance and recreation of spots, as is termed 'Teleportation' by Nasuno.

\item
Wandering motion of spots.

\end{itemize}

\paragraph{\em loop \rm}
In the experiment, at a high current region, a loop structure that moves in space is observed.
Although a connected string observed in the PW phase is similar to the loops observed in the experiment, one difference here is that in our case the strings are connected through the whole space.
Since we have chosen the periodic boundary condition, the edges of the traveling string in Fig \ref{delta-phase} are connected, to form a loop, but such traveling string observed is extended to the whole space in our model.

So far, we are not yet confident if the strings in the PW phase correspond to the loops in the experiment.
As the loops are not small in contrast to localized spot structures, the influence of the boundary may be important.
A suitable choice of boundary condition should be necessary.
Instead of periodic or Neumann boundary condition, the Dirichlet boundary may be more relevant to make more accurate correspondence with the experiment.
This problem is left for the future.

\paragraph{Related studies}
\paragraph{}
There are some recent studies concerning Nasuno's experiment on gas discharge.

Static pattern of particle structure is discussed by modifying
Gray-Scott equation in \cite{kobayashi}, while its relationship with
discharge system is not so straightfoward.
In \cite{obstructed discharge}, pattern dynamics caused by local current heating of the gas are studied by using the left-branch of Paschen
curve, termed as obstructed discharge. In our study, we consider surface charge on the electrode in the obstructed discharge in Subsection \ref{Flowing charge density},  and study instability of the homogeneous steady state at the left-branch of Paschen curve.

\subsection{Discussion}

\paragraph{Novel phenomena in reaction-diffusion system}

Since our model belongs to a class of reaction-diffusion systems, many of
the pattern dynamics observed here are common with those studied
therein; formation of spots, array of spots, and strings.
Still, the molecule structure of spot cluster, cyclic process of
collapse and split of spots, wandering motion of spots are rather novel
and characteristic to the present model.
For these phenomena, inclusion of global coupling into local
reaction-diffusion equations is essential.
Systems both with local and global couplings have also been studied in
surface catalytic reactions \cite{mikhailov 1}, coupled
maps\cite{Glocal}, and so forth.
Search for a novel class of pattern dynamics by global coupling will be
interesting in the future.

\paragraph{Itinerancy over clusters with different configurations of spots}

Through split and collapse of spots, their number changes, and also
transitions between quasi-stationary states with different clusters take place.
This transition among quasi-stationary states is found to be governed by
a specific rule with regard to the increase or decrease in the spot number.
With the aid of the phase-space representation by the macroscopic
variable $h$, this process of itinerancy over different quasi-stationary
states with different spot configurations is understood as follows:

Each quasi-stationary state corresponds to a state with a different
configuration or a different number of spots, $sp_n$.
This state is also represented by a ``saddle'' point in the high
dimensional dynamical systems of the collective coordinates $h$.
There is a stable manifold to this saddle connecting from a state with a
different number of spots, while the orbit spirals out of the saddle,
corresponding to the breathing motion of a spot.
Although this low-dimensional dynamical description of quasi-stationary
state seems to be rather effective, the transition, indeed, occurs in a
high-dimensional dynamical system.
Such switching among effectively low-dimensional states within
high-dimensional dynamical system is studied as chaotic
itinerancy\cite{GCM,chaotic scenario,milnor}.
The present model gives a novel example of chaotic itinerancy, whose
mechanism has to be elucidated in future.

When the number of spots is larger than or equal to 6, a variety of transition processes appear due to drastic increase in possible configurations of spots. 
As the spot number increases beyond 6, the combinatorial explosion of the cofiguration of spots sets in, so that stable manifolds connecting many saddle points come to be entangled.
This leads to a variety of transitions over quasi-stable spot states, resulting in chaotic itinerancy.

In \cite{magic7_1,magic7_2} it is shown that combinatorial explosion which appears beyond the degrees of freedom $5\sim 9$ leads to complex dynamics behavior with chaotic itinerancy.  Complex dynamics for the spot number beyond 6 may be discussed from this viewpoint.

\paragraph{Acknowledge}

We would like to thank Masaki Sano, Masashii Tachikawa, Akinori Awazu, Koichi Fujimoto, Shin'ichi Sasa, and Teruhisa Komatsu, for discussions and suggestions.
The present paper is dedicated to the memory of the late Dr. Satoru Nasuno.

\appendix
\section*{Appendix \normalsize \hspace{0.1in} Derivation of evolution equations }
Here we derive 2 component RD equation in terms of current density $i$, charge density $q^{\prime}(=q+s)$, from (\ref{model-1})$\sim$(\ref{model-V}).  

First, (\ref{model-1})$\sim$(\ref{model-V}) are written as
\begin{gather}
J_{k}-j_{k}+[-\frac{1}{RC}\{(q_{k}+s_{k})-(q_{k-1}+s_{k-1}) \tag*{}\\
\label{eq1} +(q_{k}+s_{k})-(q_{k+1}+s_{k+1})\}]=0 \tag{1} \\
\label{eq2}
j_{k}-(i_{k}+\frac{dq_{k}}{dt})=\frac{ds_{k}}{dt} \tag{2} \\
\label{eq3}
l\frac{di_{k}}{dt}+v_{k}-\frac{q_{k}}{C}=0 \tag{3} \\
\label{eq4}
I=\sum_{k=1}^{N}J_{k}=const. \tag{4} \\
\label{V}
V=\frac{q_{k}}{C}+\frac{s_{k}}{C}+rJ_{k}=const. \tag{5}
\end{gather}

Now, from (\ref{V}), we get
\begin{gather}
\nonumber
NV=r\sum_{k=1}^{N}J_{k}+\frac{1}{C}\sum_{k=1}^{N}(q_{k}+s_{k})=const.
\end{gather}

By transformation of variables $q^{\prime}_{k}=q_{k}+s_{k}$, and $\sum_{k=1}^{N}q_{k}^{\prime}= Q^{\prime}$, 
\begin{gather}
NV=rI+\frac{Q^{\prime}}{C} \label{Qconst} \hspace{.2in} \therefore
Q^{\prime}=const.
\end{gather}
is derived.

Now we eliminate $J_{k},j_{k}$ with (\ref{eq1}),(\ref{eq2}),(\ref{V}) to derive evolutional equation in terms of $q_{k},i_{k}$.
From (\ref{eq1}),(\ref{eq2}), we obtain
\begin{gather}
j_{k}=J_{k}+\frac{1}{RC}(q^{\prime}_{k-1}+q^{\prime}_{k+1}-2q^{\prime}_{k})
\tag*{} \\
j_{k}=i_{k}+\frac{dq^{\prime}_{k}}{dt} \tag*{}.
\end{gather}
Accordingly,
\begin{equation}
\label{eq6}
\frac{dq^{\prime}_{k}}{dt}=J_{k}-i_{k}+\frac{1}{RC}(q^{\prime}_{k-1}+q^{\prime}_{k+1}-2q^{\prime}_{k})
\end{equation}
is obtained.  
From $ J_{k}=\frac{V}{r}-\frac{q^{\prime}_{k}}{rC}$ and $v_{k}=v(i_{k})$ in Fig.\ref{nullcline}(a), (\ref{eq3}) and (\ref{eq6}) are written as
\begin{gather}
\label{eq8}
\frac{dq^{\prime}_{k}}{dt}=\frac{V}{r}-\frac{q^{\prime}_{k}}{rC}-i_{k}+\frac{1}{RC}(q^{\prime}_{k-1}+q^{\prime}_{k+1}-2q^{\prime}_{k})
\\
\label{eq9}
\frac{di_{k}}{dt}=\frac{q^{\prime}_{k}}{lC}-\frac{s_{k}}{lC}-\frac{v(i_{k})}{l},
\end{gather}
while from (\ref{eq4}), (\ref{Qconst}) and (\ref{eq6}), we get
\begin{gather}
I=\sum_{k=1}^{N}J_{k} \tag*{} \\ 
=\sum_{k=1}^{N}\{\frac{dq^{\prime}_{k}}{dt}+i_{k}-\frac{1}{RC}(q^{\prime}_{k-1}+q^{\prime}_{k+1}-2q^{\prime}_{k})\} \tag*{}
\\
=\sum_{k=1}^{N}\{i_{k}-\frac{1}{RC}(q^{\prime}_{k-1}+q^{\prime}_{k+1}-2q^{\prime}_{k})\}.\tag*{}
\end{gather}
When the boundary condition is Neumann or period one, $I=\sum_{k=1}^{N}i_{k}=const.$ is satisfied.
Therefore, from (\ref{eq9}),
\begin{gather}
\nonumber
\sum_{k=1}^{N} \frac{di_{k}}{dt}=\sum_{k=1}^{N}(
\frac{q^{\prime}_{k}}{lC}-\frac{s_{k}}{lC}-\frac{v(i_{k})}{l} ) \\
\label{eq10}
\Rightarrow\hspace{.2in}Q^{\prime}=\sum_{k=1}^{N}s_{k}+C\sum_{k=1}^{N}v(i_{k})=const.
\end{gather}
is obtained.
Since we have defined $s_{k}$ so that it vanishes when the value of each variables is spatially homogeneous, (\ref{eq10}) is rewritten as
\begin{gather}
\nonumber
Q^{\prime}=C\sum_{k=1}^{N}v(\frac{I}{N})=Cv(\frac{I}{N})N=const.
\end{gather}
Hence $Q^{\prime},V$ is determined by $\frac{I}{N}$,
\begin{gather}
\nonumber
Q^{\prime}=Cv(\frac{I}{N})N=const. \\
\nonumber
V=r\frac{I}{N}+\frac{1}{C}\frac{Q^{\prime}}{N}=r\frac{I}{N}+v(\frac{I}{N})=const.
\end{gather}
\paragraph{}

Here we discuss $s_{k}$ that is defined as spatial charge distribution.
According to Nasuno's experiment, macroscopic current-voltage characteristic $V=V(I)$ is monotonous, therefore $I$ and $V$ are kept constant, whereas local current and charge density between electrodes are not so but can show spatiotemporal patterns.
Hence we assume that local characteristic $v_{k}=v(i_{k})$ has negative slope, as is typical in glow discharge. 
Now, if there were not $s_{k}$, we got $Q^{\prime}=C\sum_{k=1}^{N}v(i_{k})=const.$ 
In homogeneous $(i_{k},q^{\prime}_{k})=(\frac{I}{N},\frac{Q^{\prime}}{N})$ case,  (\ref{eq8}), (\ref{eq9}) actually satisfied $Q^{\prime}=C\sum_{k=1}^{N}v(i_{k})(=C\sum_{k=1}^{N}v(\frac{I}{N})=Cq^{\prime}_{k}N)$ without $s_{k}$.
But, in general, (\ref{eq8}), (\ref{eq9}) can't satisfy it without $s_{k}$ in inhomogeneous $(i_{k},q^{\prime}_{k})$ case, and then need another term with physical meaning to satisfy it.
Hence, we assume that some spatial charge distributes \em besides \rm $q_{k}$.
This is why we introduce $s_{k}$.

If $s_{k}$ were independent variables, there would exist 5$N$ valuables for 4$N$ equations.
To solve the above equations, we need another constraint for the description of $s_{k}$ as a function F of $\{i_{m},q_{m}\},(m=1,2,...,N)$
\[s_{k}=F(\{i_{m},q_{m}\})\hspace{.1in} (m=1,2,...,N) \nonumber \]

This means that we assume each $s_{k}$ reaches steady state quickly, that is, $s_{k}$ is eliminated adiabatically.
This assumption appears valid, because $s_{k}$ is charge density flowing in the longitudinal direction, and therefore the time scale of it is considered to be much faster than that of charge modulation in the perpendicular direction so that the former is separated from the latter, in the case of low resistive limit of electrodes (small but non-zero $r$).

\paragraph{}
Hence $s_{k}$ are determined so as to satisfy
\begin{gather}
\frac{dq^{\prime}_{k}}{dt}=\frac{V}{r}-\frac{q^{\prime}_{k}}{rC}-i_{k}+\frac{1}{RC}(q^{\prime}_{k-1}+q^{\prime}_{k+1}-2q^{\prime}_{k})
\tag*{} \\
\frac{di_{k}}{dt}=\frac{q^{\prime}_{k}}{lC}-\frac{s_{k}}{lC}-\frac{v(i_{k})}{l}
\tag*{} \\
I=\sum_{k=1}^{N}i_{k}=const.\tag*{} \hspace{.2in} \\
Q^{\prime}=\sum_{k=1}^{N}q^{\prime}_{k}=Cv(\frac{I}{N})N=const.\tag*{} \\
V=r\frac{I}{N}+v(\frac{I}{N})=const. \tag*{}
\end{gather}

Now $s_{k}$ is written as
\[s_{k}\equiv\frac{i_{k}}{I}(Q^{\prime}-C\sum_{k=1}^{N}v(i_{k})).\]
As understood easily, with this form of $s_{k}$, it vanishes at the initial homogeneous state.
By defining $s_{k}$ in this way, our equations are solvable.
Considering spatial continuum limit and pressure effect as diffusion of $i$, the equations are described as

\begin{subequations}
\begin{gather}
\label{eq15}
\frac{dq^{\prime}}{dt}=\frac{V}{r}-\frac{q^{\prime}}{rC}-i+\frac{1}{RC}\triangle
q^{\prime}\\
\label{eq16}
\frac{di}{dt}=\frac{q^{\prime}}{lC}-\frac{s}{lC}-\frac{v(i)}{l}+D_{i}\triangle
i \\
\label{eq17}
s=\frac{i}{I}\{Q^{\prime}-C\int v(i) dS\} \\
I=\int i dS =const. \hspace{.2in} \\
Q^{\prime}=\int q^{\prime} dS =Cv(\frac{I}{S})S=const. \\
V=r\frac{I}{S}+v(\frac{I}{S})=const.
\end{gather}
\end{subequations}


\begin{thebibliography}{99}
\bibitem{footnote 2}
Actually, this logarithmic characteristic is not essential but cubic characteristic only is enough for reproducing the phenomena.
\bibitem{icvc}
In normalizing equations (\ref{model-4})$\sim$(\ref{model-9}), we determined ($i_{c},v_{c}$) so that the critical point equals to (1,1).
\bibitem{footnote 3}
Here we display only the states generated by splitting, while complicated process involving collapse of spots is not included.
\end{thebibliography}
\end{document}